\documentclass[a4paper]{article}
\def\be{\begin{equation}}
\def\eea{\end{eqnarray}}
\def\bea{\begin{eqnarray}}
\def\ee{\end{equation}}
\def\a{\alpha}
\def\b{\beta}
\def\g{\gamma}
\def\s{\sigma}
\usepackage[psamsfonts]{amssymb}
\usepackage{amsmath}
\author{M. Alimohammadi\footnote{alimohmd@ut.ac.ir}  and  Y. Naimi
\\ {\small Department of Physics, University of Tehran,}
\\ {\small North Karegar Ave., Tehran, Iran.}}
\title{ Multi-species extension of the solvable partially asymmetric reaction-diffusion
processes}
\begin{document}
\maketitle
\begin{abstract}
\noindent By considering the master equation of the partially
asymmetric diffusion process on a one-dimensional lattice, the
most general boundary condition (i.e. interactions) for the
multi-species reaction-diffusion processes is considered.
Resulting system has various interactions including diffusion to
left and right, two-particle interactions $A_\alpha A_\beta
\rightarrow A_\gamma A_\delta$ and the extended $n$-particle
drop-push interactions to left and right. We obtain three distinct
new models. The conditions on reaction rates to ensure the
solvability of the resulting models are obtained. The two-particle
conditional probabilities are calculated exactly.
\end{abstract}
\section{Introduction}
The understanding of non-equilibrium statistical physics is still
much more incomplete than that of equilibrium theory, due to the
absence of an analogue of the Boltzman-Gibbs approach and in spite
of considerable recent progress \cite{1}. Therefore
non-equilibrium systems have to be specified by some defining
dynamical rules which are then analyzed. The topic has received a
lot of attention and many reviews exist, e.g. \cite{2}-\cite{7}.

One of the interesting and important examples of the
non-equilibrium systems is the one-dimensional reaction-diffusion
processes, which have application in various fields of physics
like study of the shocks \cite{8}, noisy Burgers equation
\cite{9}, polymers in random media \cite{10}, traffic models
\cite{11}, and biopolymerization \cite{12}. As these systems are
interacting systems with $N$-particle, even simple models may pose
a formidable problem if one wants to approach them analytically.
See \cite{n1}-\cite{n4} for more recent references.

The simplest reaction-diffusion process is the totally asymmetric
simple exclusion process (TASEP). In this model, each lattice site
is occupied by at most one particle and all particles can only hop
with equal rate to their right-neighboring sites, if these sites
are not occupied. TASEP has been studied in \cite{13} by
introducing a master equation which describes the evolution
equation of the particles when they are not in neighboring sites,
and a so-called boundary condition, which specifies the situation
in which the probabilities go outside the physical regions. This
happens when some of the particles are in adjacent sites and the
master equation can not be applied to them. It has been shown that
the model is integrable in the sense that the $N$-particle
$S$-matrix is factorized into a product of two-particle
S-matrices. The coordinate Bethe ansatz has been used in this
proof.

The interesting observation is that if one chooses other boundary
conditions, with the same master equation, one can in principle
introduce other interactions (besides diffusion to
right-neighboring sites), which may be integrable in the
abovementioned sense. This is what is first done in \cite{14}, in
which the so-called drop-push model has been studied by this
method. In this model the particle hops to the next right site,
even it is occupied. It can hop by pushing all the neighboring
particles to their next right sites, with a rate depending on the
number of these particles. Some other generalization of TASEP can
be found in \cite{n5}-\cite{n7}.

The generalization of one-species reaction-diffusion processes to
$p$-species is an important task. The main problem in this
generalization, besides introducing a set of suitable boundary
conditions to model an interacting system, arises from the above
mentioned factorization of $N$-particle scattering matrix. It was
shown in \cite{15} that in order that a more-than-one species
system be solvable, in the sense of the Bethe ansatz, certain
relations should be satisfied between the rates. These relations
can be written as some kind of a spectral Yang-Baxter (SYB)
equation. By this method, all the solvable two-species
reaction-diffusion models, without annihilation and creation
reactions and with equal reaction rates, have been obtained in
\cite{15}.

The multi-species generalization of the reactions considered in
\cite{15} has been studied in \cite{16}, and the drop-push
reaction of \cite{14} has been generalized to $p$-species in
\cite{17}. The most general totally asymmetric reaction-diffusion
processes has been recently studied in \cite{18}. These processes
are \bea\label{1}
  A_\alpha\emptyset &\rightarrow &\emptyset A_\alpha \
\ \ \ {\rm with
 \ rate}\ \ D_R,\cr
  A_\alpha A_\beta &\rightarrow &A_\gamma A_\delta  \ \ \ \ {\rm with
  \ rate} \ \ c^{\alpha\beta}_{\gamma\delta},\cr
  A_\alpha A_\beta\emptyset &\rightarrow &\emptyset A_\gamma A_\delta  \ \ \ \ {\rm with
  \ rate} \ \ b^{\alpha\beta}_{\gamma\delta},\cr
   &\vdots &
    \eea
where the dots indicate the other drop-push reactions with
$n$-adjacent particles, in which in the meantime the types of the
particles can also be changed. These latter reactions are called
the extended drop-push processes. It has been shown that the
reaction rates of processes (\ref{1}) must satisfy some specific
constraints, in order that we have a set of consistent evolution
equations. Also the corresponding two-particle $S$-matrices must
satisfy the SYB equation. Some classes of the solutions of these
equations have been discussed in \cite{18}.

In all of the above studies, only the totally asymmetric exclusion
processes have been considered, i.e. the particles can only
diffuse to their next right neighboring sites. If one wants to
consider the left and right diffusions simultaneously, one must
consider a more general master equation with suitable boundary
conditions and then seek the situations in which the model is
integrable. In \cite{13}, one-species model with only simple
diffusion to left and right (i.e. partially asymmetric) has been
considered, and in \cite{19}, the one-species partially asymmetric
drop-push model has been studied. Finally a two-species model in
which the particles, besides diffusion to left and right, have
exchange-reaction has been studied in \cite{20}.

In this paper we want to study the most general $p$-species
integrable models with partially asymmetric reaction-diffusion
processes, which all the previous studied models are the special
cases of them. These general models may have some or all of the
following reactions:
 \be\label{2}
 A_\alpha\emptyset \rightarrow \emptyset A_\alpha \ \ \ \ {\rm with
 \ rate}\ \ D_R,
 \end{equation}
\be\label{3} \emptyset A_\alpha\rightarrow A_\alpha\emptyset \ \ \
\ {\rm with
 \ rate}\ \ D_L,
 \end{equation}
\be\label{4} A_\alpha A_\beta \rightarrow A_\gamma A_\delta \ \ \
\ {\rm with
 \ rate}\ \ E^{\alpha\beta}_{\gamma\delta},
 \end{equation}
 \bea\label{5}
 A_\alpha A_\beta\emptyset
&\rightarrow &\emptyset A_\gamma A_\delta  \ \ \ \ {\rm with \
rate} \ \ R^{\alpha\beta}_{\gamma\delta},\cr &\vdots &
 \eea
and
 \bea\label{6}
 \emptyset A_\alpha A_\beta &\rightarrow &A_\gamma
A_\delta\emptyset \ \ \ \ {\rm with
  \ rate} \ \ L^{\alpha\beta}_{\gamma\delta},\cr
&\vdots &
     \eea
In above equations $\a,\b,\cdots =(1,\cdots,p), \emptyset$ stands
for vacancy, and dots in eqs.(\ref{5}) and (\ref{6}) indicate the
drop-push of $n$-adjacent particles to right and left sites,
respectively, in which in the meantime the types of the particles
can also be changed. We call interactions (\ref{5}) and (\ref{6})
as right-drop-pushing and left-drop-pushing, respectively. We show
that there are three distinct models which are integrable and each
of these models contains reactions (\ref{2}) and (\ref{3}) and one
or two of the reactions (\ref{4}) to (\ref{6}).

The sheme of the paper is as follows. There are two kinds of
boundary conditions that can be generalized to $p$-species cases.
In section 2, we generalize the first kind of boundary condition,
which was introduced in  \cite{19}, to the most general
$p$-species case. Using the law of conservation of number of
particles, it is shown that there exists five constraints that
must be satisfied by reaction rates of eqs.(\ref{2})-(\ref{6}), in
order to have a set of consistent evolution equations to express
the interactions (\ref{2})-(\ref{6}). But it is seen that there is
no solution for these constraints. The situation does not change
even if we relax one of the constraints by including the
annihilation processes. Therefore one can not explain all the
reactions (\ref{2})-(\ref{6}) by this method. But it will be shown
that we can have two distinct models. In the first type model the
reactions are eqs.(\ref{2}), (\ref{3}), (\ref{4}) and (\ref{5})
and in the second type the reactions are eqs.(\ref{2}), (\ref{3}),
(\ref{4}) and (\ref{6}).

The second kind of boundary condition, which was used in \cite{13}
and \cite{20}, is generalized to the most general $p$-species case
in section 3. We show that the resulting consistent boundary
condition can explain the reactions (\ref{2}), (\ref{3}) and
(\ref{4}). This is the type 3 model. It must be mentioned that the
type 3 model is not a subclass of types 1 and 2 and is a new
distinct one. In section 4 we investigate the Bethe ansatz
solution for these models and discuss the solutions of the
corresponding SYB equations. We see that the $S$-matrix of type 3
model is much more involved than two other ones and therefore only
some special classes of solutions of its SYB equation  can be
obtained. Finally we study the conditional probabilities of these
models and in special two-particle sector, we obtain the exact
expressions.
\section{First kind generalization}
Consider a $p$-species system with particles $A_1,A_2,\cdots
,A_p$.
 The basic objects we are interested in are the probabilities
$P_{\alpha _1\cdots\alpha _N}(x_1,\cdots ,x_N;t)$ for finding at
time $t$ the particle of type $\alpha _1$ at site $x_1$, particle
of type $\alpha _2$ at site $x_2$, etc.. We take the physical
region of coordinates as $x_1<x_2<...<x_N$. The  master equation
for a partially asymmetric exclusion process is
 \bea\label{7}
 {\partial\over{\partial t}} P_{\alpha _1\cdots\alpha_N} (x_1,\cdots ,x_N;t)
 &=&D_R\sum_{i=1}^NP_{\alpha
_1\cdots\alpha_N}(x_1,\cdots,x_{i-1},x_i-1,x_{i+1},\cdots,x_N;t)\cr
 &&+
 D_L\sum_{i=1}^NP_{\alpha
 _1\cdots\alpha_N}(x_1,\cdots,x_{i-1},x_i+1,x_{i+1},\cdots,x_N;t)\cr &&-NP_{\alpha
_1\cdots\alpha_N} (x_1,\cdots ,x_N;t).
   \eea
 This equation describes
a collection of $N$ particles, diffusing to the next-right sites
by rate $D_R$ and to the next-left sites by rate $D_L$. In
eq.(\ref{7}) we have used a time scale so that
 \be\label{8}
 D_R+D_L\equiv 1.
\end{equation}
This master equation is only valid for $x_i<x_{i+1}-1$. For
$x_i=x_{i+1}-1$, there will be some terms with $x_i=x_{i+1}$ in
the right hand side of eq.(\ref{7}) which are out of the physical
region. But one can assume that (\ref{7}) is valid for all the
physical region $x_i<x_{i+1}$ by imposing certain boundary
conditions for $x_i=x_{i+1}$. Different boundary condition
introduces different interactions for particles. Following the
argument which have been given in \cite{18}, it can be easily seen
that the master equation (\ref{7}) leads to following relation for
two-particle probabilities:
 \bea\label{9}
{\partial \over{\partial
 t}}\sum_{x_2}\sum_{x_1<x_2}P_{\a_1\a_2}(x_1,x_2;t)&=&
 \sum_{x}[D_R P_{\a_1\a_2}(x,x;t)
 +D_LP_{\a_1\a_2}(x+1,x+1;t)]-\sum_x P_{\a_1\a_2}(x,x+1;t) \cr
 &=&\sum_{x} P_{\a_1\a_2}(x,x;t)-\sum_{x} P_{\a_1\a_2}(x,x+1;t).
  \eea
 This equation leads us to take $P_{\a_1\a_2}(x,x;t)$ as linear
 combination of $P_{\b_1\b_2}(x,x+1;t)$ and
 $P_{\b_1\b_2}(x-1,x;t)$'s as the only choice for having a
 consistent set of evolution equations in more-than-two-particle
 sectors \cite{18}. Therefore the most general boundary condition
 is
\be\label{10}
 P_{\a_1\a_2}(x,x) =\sum_\b b^{\b_1\b_2}_{\a_1\a_2}
 P_{\b_1\b_2}(x-1,x) +\sum_\b c^{\b_1\b_2}_{\a_1\a_2}
 P_{\b_1\b_2}(x,x+1).
 \end{equation}
 $\b$ stands for $(\b_1\b_2)$ and $b$ and $c$ are $p^2\times p^2$ matrices determine the
 interactions. In the probabilities appear in eq.(\ref{10}), we have
 suppressed all the other coordinates and the time $t$ for simplicity. In
 fact $P_{\a_1\a_2}(x,x):=P_{\gamma_1\cdots\gamma_i\alpha_1\alpha_2\gamma _{i+3}\cdots\gamma
 _N}(x_1,\cdots,x_i,x,x,x_{i+3},\cdots,x_N)$.
 In the first step, let us exclude the creation and annihilation
 processes (it can be shown that one can not study the creation
 processes by this method, so in fact in this step, we exclude the
 annihilation processes). Since the number of particles is constant
 in time, summing over  $\a_1$ and $\a_2$ makes the left-hand side
 of (\ref{9}) zero and results:
 \be\label{11}
 -\sum_x\sum_\a P_{\a_1\a_2}(x,x+1)+
 \sum_x\sum_\b\left(\sum_\a(b+c)^{\b_1\b_2}_{\a_1\a_2}\right)
 P_{\b_1\b_2}(x,x+1)=0,
 \end{equation}
in which eq.(\ref{10}) has been used. Clearly eq.(\ref{11}) gives:
 \be\label{12}
 \sum_\a(b+c)^{\b_1\b_2}_{\a_1\a_2}=1 \ \ \ \ {\rm constraint\ \ (I)}.
 \end{equation}
Note that in $p=1$, the boundary condition (\ref{10}) and
constraint (\ref{12}) reduce to those considered in \cite{19}.
Also in the case of totally asymmetric processes in which $D_L=0$,
our problem reduces to one considered in \cite{18}. Following the
same steps as \cite{18}, we first consider ${\dot
P}_{\a_1\a_2}(x,x+1)$. Using eqs.(\ref{7}) and (\ref{10}), it is
found
 \bea\nonumber
 {\dot P}_{\a_1\a_2}(x,x+1)&=&
 D_RP_{\a_1\a_2}(x-1,x+1)+D_LP_{\a_1\a_2}(x,x+2)\cr
 &&+D_R\sum_\b b^{\b_1\b_2}_{\a_1\a_2}P_{\b_1\b_2}(x-1,x)
 +D_L\sum_\b c^{\b_1\b_2}_{\a_1\a_2}P_{\b_1\b_2}(x+1,x+2)\cr
 &&+(D_R\sum_\b
c^{\b_1\b_2}_{\a_1\a_2}+D_L\sum_\b
b^{\b_1\b_2}_{\a_1\a_2})P_{\b_1\b_2}(x,x+1)
 -2P_{\a_1\a_2}(x,x+1)\cr
 &=& D_RP_{\a_1\a_2}(x-1,x+1)+D_LP_{\a_1\a_2}(x,x+2)+
 D_R\sum_\b b^{\b_1\b_2}_{\a_1\a_2}P_{\b_1\b_2}(x-1,x)\cr
 &&+D_L\sum_\b c^{\b_1\b_2}_{\a_1\a_2}P_{\b_1\b_2}(x+1,x+2)+\sum_{\b\neq \a}(D_R
c^{\b_1\b_2}_{\a_1\a_2}+D_Lb^{\b_1\b_2}_{\a_1\a_2})P_{\b_1\b_2}(x,x+1)\cr
&&-[D_R+D_L+\sum_{\b\neq
\a}(D_Rc^{\a_1\a_2}_{\b_1\b_2}+D_Lb^{\a_1\a_2}_{\b_1\b_2})+\sum_{\b
}(D_Rb^{\a_1\a_2}_{\b_1\b_2}+D_Lc^{\a_1\a_2}_{\b_1\b_2})]P_{\a_1\a_2}(x,x+1),
\eea
  \be\label{13}
\end{equation}
 in which we use eqs.(\ref{8}) and (\ref{12}). The latter can be
written as: \be\label{14}
 c^{\a_1\a_2}_{\a_1\a_2}= 1- \sum_\b b_{\b_1\b_2}^{\a_1\a_2}- \sum_{\b\neq \a}
 c^{\a_1\a_2}_{\b_1\b_2},
 \end{equation}
or
 \be\label{15}
 b^{\a_1\a_2}_{\a_1\a_2}= 1- \sum_{\b\neq \a}
 b^{\a_1\a_2}_{\b_1\b_2} - \sum_\b c_{\b_1\b_2}^{\a_1\a_2}.
 \end{equation}
It is seen that the evolution equation (\ref{13}) describes the
following two-particle interactions:
 \bea\label{16}
 A_\alpha\emptyset &\rightarrow
&\emptyset A_\alpha \ \ \ \ {\rm with
 \ rate}\ \ D_R,\cr
 \emptyset A_\alpha&\rightarrow
&A_\alpha\emptyset \ \ \ \ {\rm with
 \ rate}\ \ D_L,\cr
 A_\alpha A_\beta &\rightarrow &A_\gamma
A_\delta \ \ \ \ {\rm with
 \ rate}\ \ D_Rc^{\alpha\beta}_{\gamma\delta}+D_Lb^{\alpha\beta}_{\gamma\delta},\cr
 A_\alpha A_\beta\emptyset
&\rightarrow &\emptyset A_\gamma A_\delta  \ \ \ \ {\rm with \
rate} \ \ D_Rb^{\alpha\beta}_{\gamma\delta},\cr
 \emptyset A_\alpha A_\beta &\rightarrow &A_\gamma A_\delta\emptyset \ \ \ \ {\rm with
  \ rate} \ \ D_Lc^{\alpha\beta}_{\gamma\delta}.
 \eea
To study the consistency of our formalism and also deriving the
more-than-two particle interactions, we consider${\dot
P}_{\a_1\cdots\a_n}(x,x+1,\cdots,x+n-1)$. In $n=3$, we encounter
two boundary terms $P_{\a_1\a_2\a_3}(x,x+1,x+1)$ and
$P_{\a_1\a_2\a_3}(x+1,x+1,x+2)$. Using (\ref{10}), the first one
becomes: \bea\label{17}
 P_{\a_1\a_2\a_3}(x,x+1,x+1)&=& \sum_{\b\g} b^{\b_2\b_3}_{\a_2\a_3}
[b^{\g_1\g_2}_{\a_1\b_2}
 P_{\g_1\g_2\b_3}(x-1,x,x+1)+c^{\g_1\g_2}_{\a_1\b_2}
 P_{\g_1\g_2\b_3}(x,x+1,x+1)]\cr &&+ \sum_\b c^{\b_2\b_3}_{\a_2\a_3}
 P_{\a_1\b_2\b_3}(x,x+1,x+2)
 \eea
 which describes the boundary term $P_{\a_1\a_2\a_3}(x,x+1,x+1)$ as
 a linear combination of other boundary terms, i.e. $P_{\g_1\g_2\b_3}(x,x+1,x+1)$'s. As has been shown in
 \cite{18}, the only consistent solution to this problem is the
 vanishing of these terms in the right-hand side of eq.(\ref{17}), which
 results:
\be\label{18}
 \sum_{\b_2} c^{\g_1\g_2}_{\a_1\b_2}b^{\b_2\b_3}_{\a_2\a_3}=0\ \ \ \ {\rm constraint\ \
(II) },
 \end{equation}
or
 \be\label{19}
 (1\otimes b)(c\otimes 1)=0,
 \end{equation}
 in which 1 stands for the $p\times p$ identity matrix. The second
 boundary term is
\bea\label{20}
 P_{\a_1\a_2\a_3}&(&x+1,x+1,x+2)= \sum_\b b^{\b_1\b_2}_{\a_1\a_2}
 P_{\b_1\b_2\a_3}(x,x+1,x+2)\cr
 &&+\sum_{\b\g} c^{\b_1\b_2}_{\a_1\a_2}
 [b^{\g_2\g_3}_{\b_2\a_3}
 P_{\b_1\g_2\g_3}(x+1,x+1,x+2)+c^{\g_2\g_3}_{\b_2\a_3}
 P_{\b_1\g_2\g_3}(x+1,x+2,x+3)], \ \ \ \ \ \
 \eea
which again leads us to take
 \be\label{21}
 \sum_{\b_2} c^{\b_1\b_2}_{\a_1\a_2}b^{\g_2\g_3}_{\b_2\a_3}=0\ \ \ \ {\rm constraint\ \
(III) },
\end{equation}
or \be\label{22}
 (c\otimes 1)(1\otimes b)=0.
 \end{equation}
Assuming constraints (\ref{18}) and (\ref{21}) and using
eqs.(\ref{7}) and (\ref{10}),  ${\dot P}_{\vec
 \a}(x,x+1,x+2)$ is
 \bea\nonumber
 {\dot P}_{\vec
 \a}(x,x+1,x+2)&=&
 D_RP_{\vec
 \a}(x-1,x+1,x+2)+D_LP_{\vec
 \a}(x,x+1,x+3)\cr&&+
 D_R\sum_\b b^{\b_1\b_2}_{\a_1\a_2}P_{\b_1\b_2\a_3}(x-1,x,x+2)
 +D_L\sum_\b c^{\b_2\b_3}_{\a_2\a_3}P_{\a_1\b_2\b_3}(x,x+2,x+3) \cr  &&
 +\sum_{\b\neq \a}(D_R
c^{\b_1\b_2}_{\a_1\a_2}+D_Lb^{\b_1\b_2}_{\a_1\a_2})P_{\b_1\b_2\a_3}(x,x+1,x+2)\cr
&&+\sum_{\b\neq \a}(D_R
c^{\b_2\b_3}_{\a_2\a_3}+D_Lb^{\b_2\b_3}_{\a_2\a_3})P_{\a_1\b_2\b_3}(x,x+1,x+2)\cr
&&+D_R\sum_\g b^{\vec\g}_{\vec\a}P_{\vec \g}(x-1,x,x+1)
+D_L\sum_\g c^{\vec\g}_{\vec\a}P_{\vec \g}(x+1,x+2,x+3)\cr
&&-[D_R+D_L +D_R\sum_\b b^{\a_1\a_2}_{\b_1\b_2} +D_L\sum_\b
c^{\a_1\a_2}_{\b_1\b_2}+D_R\sum_\b
b^{\a_2\a_3}_{\b_2\b_3}+D_L\sum_\b c^{\a_2\a_3}_{\b_2\b_3} \cr
&&+\sum_{\b\neq \a}(D_R
c^{\a_1\a_2}_{\b_1\b_2}+D_Lb^{\a_1\a_2}_{\b_1\b_2})+\sum_{\b\neq
\a}(D_Rc^{\a_2\a_3}_{\b_2\b_3}+D_Lb^{\a_2\a_3}_{\b_2\b_3})
]P_{\vec
 \a}(x,x+1,x+2),
\eea \be\label{23}
\end{equation}
 in which we have used eqs.(\ref{14}) and (\ref{15}) for
diagonal elements of matrix $D_Rc+D_Lb$. $b^{\vec\g}_{\vec\a}$ and
$c^{\vec\g}_{\vec\a}$ are defined as following
 \be\label{24}
 b^{\vec\g}_{\vec\a}=\sum_\g b^{\g_1\g_2}_{\a_1\b}
 b^{\b\g_3}_{\a_2\a_3},
 \end{equation}
 \be\label{25}
 c^{\vec\g}_{\vec\a}=\sum_\g c^{\g_1\b}_{\a_1\a_2}
 c^{\g_2\g_3}_{\b\a_3}.
 \end{equation}
Looking at source terms of eq.(\ref{23}), it is obvious that they
describe the reactions (\ref{16}) and the following three-particle
drop-push reactions: \be\label{26}
 A_{\g_1}A_{\g_2}A_{\g_3}\emptyset \rightarrow\emptyset A_{\a_1}A_{\a_2}A_{\a_3} \ \ \ \ {\rm with
  \ rate} \ \ D_R b^{\vec\g}_{\vec\a},
\end{equation}
and
 \be\label{27}
 \emptyset A_{\g_1}A_{\g_2}A_{\g_3} \rightarrow A_{\a_1}A_{\a_2}A_{\a_3}\emptyset \ \ \ \ {\rm with
  \ rate} \ \ D_Lc^{\vec\g}_{\vec\a}.
 \end{equation}
The sink terms are consistent with this description, provided
\be\label{28}
 \sum_\b b^{\vec\a}_{\vec\b}=\sum_{\b\g} b^{\a_1\a_2}_{\b_1\g}
 b^{\g\a_3}_{\b_2\b_3}=\sum_\b b^{\a_1\a_2}_{\b_1\b_2}\ \ \ \ {\rm
 constraint \ \ (IV) },
 \end{equation}
and \be\label{29}
 \sum_\b c^{\vec\a}_{\vec\b}=\sum_{\b\g} c^{\a_1\g}_{\b_1\b_2}
 c^{\a_2\a_3}_{\g\b_3}=\sum_\b c^{\a_2\a_3}_{\b_2\b_3}\ \ \ \ {\rm constraint \ \ (V) }.
 \end{equation}
By calculating other ${\dot P}_{\vec \a}(x,x+1,\cdots,x+n-1)$'s it
can be shown that we need not any more constraints and therefore
the master equation (\ref{7}) with boundary condition (\ref{10})
and five constraints (I)-(V) can consistently describe the
following reactions:

  \be\label{30}
 A_\alpha\emptyset \rightarrow \emptyset A_\alpha \ \ \ \ {\rm with
 \ rate}\ \ D_R
\end{equation}
\be\label{31}
 \emptyset A_\alpha\rightarrow A_\alpha\emptyset \
\ \ \ {\rm with
 \ rate}\ \ D_L
\end{equation}
 \be\label{32}
  A_\alpha A_\beta \rightarrow A_\gamma A_\delta  \
\ \ \ {\rm with
  \ rate} \ \ D_R c^{\alpha\beta}_{\gamma\delta}+D_L b^{\alpha\beta}_{\gamma\delta}
\end{equation}
\be\label{33} A_{\a_0}\cdots A_{\a_n}\emptyset \rightarrow
\emptyset A_{\g_0}\cdots A_{\g_n}  \ \ \ \ {\rm with
  \ rate} \ \ D_R(b_{n-1,n}\cdots
  b_{0,1})^{\a_0\cdots\a_n}_{\g_0\cdots\g_n},
\end{equation}
\be\label{34} \emptyset A_{\a_0}\cdots A_{\a_n} \rightarrow
A_{\g_0}\cdots A_{\g_n}\emptyset  \ \ \ \ {\rm with
  \ rate} \ \ D_L(c_{0,1}\cdots
  c_{n-1,n})^{\a_0\cdots\a_n}_{\g_0\cdots\g_n}.
\end{equation}
In above equations we use the following definition for $b_{k,k+1}$
and $c_{k,k+1}$ : \be\label{35}
 a_{k,k+1}=1\otimes\cdots\otimes
 1\otimes\underbrace{a}_{k,k+1}\otimes 1 \otimes\cdots \otimes 1.
 \end{equation}
Note that for $D_L=0$, the five classes of the above reactions
reduce to three ones discussed in  \cite{18}. In \cite{18}, the
constraints between reaction rates are three relations (I), (II),
and (IV). Note that at $D_L=0$, the constraints (III) and (V) do
not appear since the multiplication factors of their corresponding
terms in evolution equation is $D_L$ , which is zero.

To find the set of solutions of five constraints (I) to (V), one
can consider the solutions of equations (I), (II) and (IV), that
is the solutions derived in \cite{18}, and then considers the
subset of them satisfys (III) and (V). We must note that in our
models, the diagonal elements of matrix $c$ are the reaction rates
of the last line of eq.(\ref{16}) and must be positive. This is in
contrast to the case studied in \cite{18} in which the diagonal
elements of $c$ can be negative.

We can also follow another approach. That is trying to find the
solution of equations (II) to (IV) and then seek ones which
satisfy relation (I). As these relations are rather complex, we
can not completely solve them for arbitrary $p$, but we try them
as much as possible.

As all the matrix elements of matrices $b$ and $c$ are reaction
rates, they can not be negative, so the only solution of
eq.(\ref{18}) is:
 \be\label{36}
c^{\g_1\g_2}_{\a_1\b_2}b^{\b_2\b_3}_{\a_2\a_3}=0 \ \ \ \ {\rm
(without \ sum \ over }\ \b_2).
 \end{equation}
This relation has two following solutions ( for each $\b_2$)
\be\label{37}
 c^{\g_1\g_2}_{\a_1\b_2}=0\ \ \ \  {\rm and}  \ \ \ \
 b^{\b_2\b_3}_{\a_2\a_3}=0.
 \end{equation}
So for  each $\b_2$ we have two solutions, and as $\b_2$ runs from
1 to $p $, we have $2^p-2$ set of solutions for constraint (II).
We exclude two of the solutions in which all of the elements of
$c$ or $b$ is zero , since we look for the situations in which
$b\neq 0$ and $c\neq 0$. We will later study the cases $b=0$ or
$c=0$ in which the number of independent classes of reactions
(\ref{30})-(\ref{34}) reduces to four. By the same argument, the
solutions of eq.(\ref{21}) are (for each $\b_2$)
 \be\label{38}
 c^{\b_1\b_2}_{\a_1\a_2}=0\ \ \ \ {\rm and} \ \ \ \ b^{\g_2\g_3}_{\b_2\a_3}=0,
 \end{equation}
and therefore we again have $2^p-2$ set of solutions for
constraint (III). Note that from $2^p-2$ solutions of constraints
II ( and III ) only ($p-1$) of them are independent, that is does
not transform to each other under interchanging of the labels of
the species of the particles. So the number of independent
solutions of constraints (II) and (III) are $(p-1)(2^p-2)$. For
example in $p=2$, the independent solutions of (II) and (III) are
\bea\label{39}
 &\{&c^{\g_1\g_2}_{\a_11}=0 , b^{2\b_3}_{\a_2\a_3}=0
 ,c^{\b_11}_{\a_1\a_2}=0 , b^{\g_2\g_3}_{2\a_3}=0\},\cr
 &\{&c^{\g_1\g_2}_{\a_11}=0 , b^{2\b_3}_{\a_2\a_3}=0
 ,c^{\b_12}_{\a_1\a_2}=0 , b^{\g_2\g_3}_{1\a_3}=0\},
 \eea
which can be written as
 \be\label{40}
 b=\left(
 \begin{array}{cccc}
  b_{11}&b_{12}& 0 & 0 \\
  b_{21}&b_{22}&0&0 \\
 0 & 0 &0&0\\
 0 & 0 & 0 &0
 \end{array}
 \right) \ , \
 c=\left(
 \begin{array}{cccc}
 0 &0 &0  & 0 \\
 0 &c_{22} &0 & c_{24} \\
 0 &0 &0&0\\
 0 & c_{42} & 0&c_{44}
 \end{array}
 \right),
 \end{equation}
and \be\label{41}
 b=\left(
 \begin{array}{cccc}
  0&0& 0 & 0 \\
  0&0&0&0 \\
 b_{31} & b_{32} &0&0\\
 b_{41} & b_{42} & 0 &0
 \end{array}
 \right) \ , \
 c=\left(
 \begin{array}{cccc}
 0 &0 &0  & 0 \\
 c_{21} &0 &c_{23} & 0 \\
 0 &0 &0&0\\
 c_{41} & 0 & c_{43}&0
 \end{array}
 \right),
 \end{equation}
 respectively. We label the states as $ |1>=(1,1) \ , \ |2>=(1,2) \ , \ |3>=(2,1) \ {\rm and}  \
 |4>=(2,2)$. Putting eqs.(\ref{40}) and (\ref{41}) into the
 constraint (IV) and (V) (eqs.(\ref{28}) and (\ref{29})) results
\be\label{42}
 b=\left(
 \begin{array}{cccc}
  1&1& 0 & 0 \\
  b_{21}&b_{21}&0&0 \\
 0 & 0 &0&0\\
 0 & 0 & 0 &0
 \end{array}
 \right) \ , \
 c=\left(
 \begin{array}{cccc}
 0 &0 &0  & 0 \\
 0 &c_{22} &0 & c_{22} \\
 0 &0 &0&0\\
 0 & 1 & 0&1
 \end{array}
 \right),
 \end{equation}
and \be\label{43}
 b=\left(
 \begin{array}{cccc}
  0&0& 0 & 0 \\
  0&0&0&0 \\
 1 &1 &0&0\\
 b_{41} & b_{41} & 0 &0
 \end{array}
 \right) \ , \
 c=\left(
 \begin{array}{cccc}
 0 &0 &0  & 0 \\
 1 &0 &1 & 0 \\
 0 &0 &0&0\\
 c_{41} & 0 & c_{41}&0
 \end{array}
 \right),
 \end{equation}
 respectively. Now eq.(\ref{12}) (constraint I) says that the sum
 of the elements of each column of matrix ($b+c$) must be one, which
 unfortunately does not satisfy by (\ref{42}) and (\ref{43}). The
 sum of the elements of the second column of ($b+c$) of eq.(\ref{42}) and
 the first column of eq.(\ref{43}) are greater than (or equal to) 2. So
 reactions (\ref{30})-(\ref{34}) have not any representation in
 $p=2$, the situation which we expect to be true for other $p$'s.
 For example in $p=3$, constraints (II) and (III) have 12
 independent solutions, which are in two categories: the number of
 constraints on $b_i^{\ j}$ and $c_k^{\ l}$'s are equal (three on $b_i^{\ j}$ and three on $c_k^{\
 l}$), and one which these numbers differ, i.e. 4 and 2.
As an example of the first category, we consider the case in which
 $c_{\a1}^{\ \ \ j}=0, c_{\a2}^{\ \ \ j}=0, c_i^{\ \b 1}=0 , b_k^{\ 3\g}=0, b_{1\a}^{\ \ \ k}=0,
 b_{3\a}^{\ \ \  k}=0$. It means that in matrix $c$, the rows 1,2,4,5,7,8 and
 columns 2,5, and 8 are zero, so it has 18 non-zero elements, and
 $b$ is a matrix in which the rows 1,2,3,7,8,9 and columns 7,8, and
 9 are zero so it also has 18 non-zero elements. Putting these $b$
and $c$ matrices in constraint (IV) and (V) results two following
solutions for each $b$ and $c$:
  \bea\label{44}
c_1:\{c_{31}=c_{34}=c_{37}=1, c_{33}  \ \ {\rm and}  \ \ c_{36}\ \
{\rm \  arbitrary} \},\cr c_2:\{c_{93}=c_{96}=c_{99}=1, c_{91} \ \
{\rm and} \ \ c_{94}\ \ {\rm \ arbitrary} \},\cr
 b_1:\{b_{41}=b_{42}=b_{43}=1,b_{44}
\ \ {\rm and}\ \ b_{45}\ \  {\rm \ arbitrary} \},\cr
b_2:\{b_{54}=b_{55}=b_{56}=1,b_{51} \ \ {\rm and}\ \ b_{52}\ \
{\rm \ arbitrary} \},
  \eea
in which we only write down the non-zero elements. It can be
easily seen that none of the combinations $b_1+c_1$, $b_1+c_2$,
$b_2+c_1$, $b_2+c_2$ are acceptable in the sense of constraint
(I), as at least the sum of the elements of one of the columns of
these matrices are greater than (or equal to) 2. We have checked
that the same situation arises in other 11 solutions. So again in
$p=3$, we have no representation. We can not generally prove this,
but we believe that the set of constraints (I)-(V) have no
solution for arbitrary $p$.

One may suppose that if we somehow change the constraint (I), then
it may be possible to find some solution for our equations. So we
add the annihilation processes to our previous interactions. Note
that these interactions appear only in the sink terms of the
evolution equation, as if we consider the initial state with $n$
particles,
 no annihilation processes can lead to a $n$-particle state at any
other time $t$. So if we change the constraint (I) to: (as we have
not the conservation of particles)
 \be\label{45}
 \sum_\a(b+c)^{\b_1\b_2}_{\a_1\a_2}=1-\lambda_{\b_1\b_2},
 \end{equation}
and using it in calculation of ${\dot P}_{\a_1\a_2}(x,x+1)$, we
find the same equation as (\ref{13}), except an extra term
$\lambda_{\a_1\a_2}{\dot P}_{\a_1\a_2}(x,x+1)$ which is added to
sink terms. So $\lambda_{\a_1\a_2}$ is the sum of the rates of all
annihilation processes with initial state ($\a_1\a_2$) and
therefore is a positive quantity. Therefore adding the
annihilation processes to interactions (\ref{30})-(\ref{34}) means
that the sum of the elements of each column of ($b+c$) can now be
less than or equal to one. But as we have shown in eqs.(\ref{42}),
(\ref{43}) and (\ref{44}), the sum of the elements of some of the
columns of ($b+c$) in these examples are at least 2, which differs
from what is suggested by eq.(\ref{45}). In brief, including the
annihilation processes can not alter our result and the set of
processes (\ref{30})-(\ref{34}) have no representation, with or
without adding the annihilation processes.

Now it is interesting to note that even if one of the matrices $b$
or $c$ be equal to zero, we have yet all four desired reactions:
Diffusion to left {\it and} right, two-particle reactions
$A_\alpha A_\beta\rightarrow A_\gamma A_\delta$, and the extended
drop-push reactions, which the latter occur only in one side (left
{\it or} right). These are almost the general reactions that one
can study in this framework. Let us check the constraints in these
cases.
\subsection{Type 1 model}
Take $c=0$. Eq.(12) becomes
 \be\label{46}
 \sum_\a b^{\b_1\b_2}_{\a_1\a_2}=1 .
 \end{equation}
Constraints (II), (III) and (V) are satisfied trivially and
constraint (IV) is also satisfied: using eq.(\ref{46}), both sides
of (IV) become one. Therefore master equation (\ref{7}) with
boundary condition
 \be\label{47}
 P_{\a_1\a_2}(x,x) =\sum_\b b^{\b_1\b_2}_{\a_1\a_2}
 P_{\b_1\b_2}(x-1,x),
\end{equation}
and constraint (\ref{46}), describe consistently the following
reactions:
 \bea\label{48}
 A_\alpha\emptyset &\rightarrow
&\emptyset A_\alpha \ \ \ \ {\rm with
 \ rate}\ \ D_R,\cr
 \emptyset A_\alpha&\rightarrow
&A_\alpha\emptyset \ \ \ \ {\rm with
 \ rate}\ \ D_L,\cr
 A_\alpha A_\beta &\rightarrow &A_\gamma A_\delta \ \ \ \ {\rm
with
 \ rate}\ \ D_L b^{\alpha\beta}_{\gamma\delta},\cr
A_{\a_0}\cdots A_{\a_n}\emptyset &\rightarrow &\emptyset
A_{\g_0}\cdots A_{\g_n}  \ \ \ \ {\rm with
  \ rate} \ \ D_R(b_{n-1,n}\cdots
  b_{0,1})^{\a_0\cdots\a_n}_{\g_0\cdots\g_n}.
    \eea
\subsection{Type 2 model}
 In the same way, for $b=0$ it can be
seen that the master equation (\ref{7}) with boundary condition
\be\label{49}
 P_{\a_1\a_2}(x,x) =\sum_\b c^{\b_1\b_2}_{\a_1\a_2}
 P_{\b_1\b_2}(x,x+1),
\end{equation}
and constraint \be\label{50}
 \sum_\a c^{\b_1\b_2}_{\a_1\a_2}=1 ,
 \end{equation}
describe successfully the reactions:
 \bea\label{51}
 A_\alpha\emptyset &\rightarrow
&\emptyset A_\alpha \ \ \ \ {\rm with
 \ rate}\ \ D_R,\cr
 \emptyset A_\alpha&\rightarrow
&A_\alpha\emptyset \ \ \ \ {\rm with
 \ rate}\ \ D_L,\cr
 A_\alpha A_\beta &\rightarrow &A_\gamma A_\delta \ \ \ \ {\rm
with
 \ rate}\ \ D_R c^{\alpha\beta}_{\gamma\delta},\cr
\emptyset A_{\a_0}\cdots A_{\a_n} &\rightarrow & A_{\g_0}\cdots
A_{\g_n}\emptyset  \ \ \ \ {\rm with
  \ rate} \ \ D_L(c_{0,1}\cdots
  c_{n-1,n})^{\a_0\cdots\a_n}_{\g_0\cdots\g_n}.
    \eea
The condition of solvability of these models will be discussed in
next sections.
\section{Second kind generalization}
By noting the first line of eq.(\ref{9}), it is seen that
eq.(\ref{10}) is not the only possible $p$-species boundary
condition. In fact, one can instead consider the following
boundary condition : \be\label{52}
 D_RP_{\a_1\a_2}(x,x) +D_LP_{\a_1\a_2}(x+1,x+1)=\sum_\b b^{\b_1\b_2}_{\a_1\a_2}
 P_{\b_1\b_2}(x-1,x)+\sum_\b c^{\b_1\b_2}_{\a_1\a_2}
 P_{\b_1\b_2}(x,x+1).
 \end{equation}
 This is the multi-species generalization of the boundary condition considered in
 \cite{13} and \cite{20}.

 To study the interactions introduced by (\ref{7}) and (\ref{52}), we
 must again consider ${\dot
P}_{\a_1\cdots\a_n}(x,x+1,\cdots,x+n-1)$. In $n=3$, we encounter
the boundary term $D_RP_{\a_1\a_2\a_3}(x,x+1,x+1)
+D_LP_{\a_1\a_2\a_3}(x,x+2,x+2)$, where using (\ref{52}) results
$\sum_\b b^{\b_2\b_3}_{\a_2\a_3}
 P_{\a_1\b_2\b_3}(x,x,x+1)+\sum_\b c^{\b_2\b_3}_{\a_2\a_3}
 P_{\a_1\b_2\b_3}(x,x+1,x+2)$. But the first term
 $P_{\a_1\b_2\b_3}(x,x,x+1)$ can not be written in terms of
 physical probabilities, since in this case only the linear
 combination $D_RP_{\a_1\a_2\cdots}(x,x,\cdots)
 +D_LP_{\a_1\a_2\cdots}(x+1,x+1,\cdots)$ can be written in terms of
 physical function (eq.(\ref{52})). This is in contrast with the
 case studied in section 2. The only solution to this problem is
 taking
\be\label{53} b=0.
\end{equation}
So our second kind $p$-species model is defined through the master
equation (\ref{7}) and the following boundary condition :
\be\label{54}
 D_RP_{\a_1\a_2}(x,x) +D_LP_{\a_1\a_2}(x+1,x+1)=\sum_\b c^{\b_1\b_2}_{\a_1\a_2}
 P_{\b_1\b_2}(x,x+1).
 \end{equation}
Conservation of number of particle gives \be\label{55}
 \sum_\a c^{\b_1\b_2}_{\a_1\a_2}=1 ,
 \end{equation}
and calculating ${\dot P}_{\a_1\a_2}(x,x+1)$ results
\bea\label{56} {\dot P}_{\a_1\a_2}(x,x+1)&=&
 D_RP_{\a_1\a_2}(x-1,x+1)+D_LP_{\a_1\a_2}(x,x+2)+
 \sum_{\b\neq
\a} c^{\b_1\b_2}_{\a_1\a_2}P_{\b_1\b_2}(x,x+1)\cr
 &&-(D_R+D_L+\sum_{\b\neq
\a} c^{\a_1\a_2}_{\b_1\b_2})P_{\a_1\a_2}(x,x+1).
 \eea
This equation describes the following reactions as source and sink
terms \bea\label{57}
 A_\alpha\emptyset &\rightarrow &\emptyset
A_\alpha \ \ \ \ {\rm with
 \ rate}\ \ D_R,\cr
 \emptyset A_\alpha&\rightarrow
&A_\alpha\emptyset \ \ \ \ {\rm with
 \ rate}\ \ D_L,\cr
 A_\alpha A_\beta &\rightarrow &A_\gamma
A_\delta \ \ \ \ {\rm with
 \ rate}\ \ c^{\alpha\beta}_{\gamma\delta}.
 \eea
 Calculating other ${\dot
P}_{\a_1\cdots\a_n}(x,x+1,\cdots,x+n-1)$'s confirms these
reactions without any further constraint. So the {\bf type 3
model} is defined by master equation (\ref{7}), boundary condition
(\ref{54}), constraint (\ref{55}), and reactions (\ref{57}).
\section{Bethe ansatz solution }
Until now, we have constructed a consistent formalism to study
some reaction-diffusion processes. Now we want to solve the
resulting evolution equations and check the solvability of these
models. To solve the master equation (\ref{7}), we consider the
following Bethe ansatz
 \begin{equation}\label{58}
P_{\alpha_1\cdots \alpha _N}({\mathbf x};t)=e^{-E_Nt}\psi
_{\alpha_1\cdots\alpha _N}({\mathbf x}),
\end{equation}
with
 \begin{equation}\label{59}
\Psi({\mathbf x})=\sum_\sigma {\mathbf{A}}_\sigma e^{i\sigma
({\mathbf{p}}).{\mathbf{x}}}.
\end{equation}
$\Psi$ is a tensor of rank $N$ with components $\psi
_{\alpha_1\cdots\alpha _N}({\mathbf x})$ and the summation runs
over the elements of the permutation group of $N$ objects
\cite{n8,n9}. Inserting (\ref{58}) in (\ref{7}), results
\be\label{60}
 E_N=\sum_{k=1}^N(1-D_Re^{-ip_k}-D_Le^{ip_k}).
 \end{equation}
Inserting (\ref{58}) in boundary condition (\ref{47}) gives
\be\label{61}
  \Psi(\cdots ,x_k=x,x_{k+1}=x,\cdots)=b_{k,k+1}
  \Psi(\cdots ,x_k=x-1,x_{k+1}=x,\cdots ),
  \end{equation}
which using (\ref{59}) results
 \be\label{62}
 [1-e^{-i\sigma (p_k)}b_{k,k+1}]
 {\mathbf A}_\sigma +
 [1-e^{-i\sigma (p_{k+1})}b_{k,k+1}]
 {\mathbf A}_{\sigma\sigma_k}=0.
 \end{equation}
 $\sigma_k$ is an element of permutation group which only
interchanges $p_k$ and $p_{k+1}$: \be\label{63}
 \s_k:(p_1,\cdots ,p_k,p_{k+1},\cdots ,p_N)\rightarrow
 (p_1,\cdots ,p_{k+1},p_{k},\cdots ,p_N).
 \end{equation}
 Eq.(\ref{62}) gives ${\mathbf A}_{\sigma\sigma_k}$ in terms of ${\mathbf
 A}_{\sigma}$ as following:
 \be\label{64}
 {\mathbf A}_{\sigma\sigma_k}=S_{k,k+1}^{(1)}(\s (p_k),\s (p_{k+1}))
 {\mathbf A}_{\sigma},
 \end{equation}
where \be\label{65}
 S_{k,k+1}^{(1)}(z_1,z_2)=1\otimes\cdots\otimes
 1\otimes\underbrace{S^{(1)}(z_1,z_2)}_{k,k+1}\otimes 1 \otimes\cdots \otimes 1,
 \end{equation}
and $S^{(1)}(z_1,z_2)$ is the following $p^2\times p^2$ matrix
\be\label{66}
 S^{(1)}(z_1,z_2)=-(1-z_2^{-1}b)^{-1}(1-z_1^{-1}b),
 \end{equation}
in which  $z_k=e^{ip_k}$. The same procedure for boundary
conditions (\ref{49}) and (\ref{54}), i.e. the type 2 and type 3
models, results
 \be\label{67}
 S^{(2)}(z_1,z_2)=-(1-z_1c)^{-1}(1-z_2c),
 \end{equation}
 and
\be\label{68}
 S^{(3)}(z_1,z_2)=-(D_R+z_1z_2D_L-z_1c)^{-1}(D_R+z_1z_2D_L-z_2c),
 \end{equation}
respectively. Eq.(\ref{64}) allows one to compute all ${\mathbf
A}_{\sigma}$'s in terms of ${\mathbf A}_{1}$(which is set to
unity).

As the generators of permutation group satisfy
$\s_k\s_{k+1}\s_k=\s_{k+1}\s_k\s_{k+1}$, so one also needs
\be\label{69}
 {\mathbf A}_{\s_k\s_{k+1}\s_k}={\mathbf A}_{\s_{k+1}\s_k\s_{k+1}}.
 \end{equation}
This, in terms of $S$-matrices becomes
  \be\label{70}
 S_{12}(z_2,z_3)S_{23}(z_1,z_3)S_{12}(z_1,z_2)=
 S_{23}(z_1,z_2)S_{12}(z_1,z_3)S_{23}(z_2,z_3).
 \end{equation}
In the terms of $R$-matrix defined through
 \be\label{71}
 S_{k,k+1}=:\Pi_{k,k+1}R_{k,k+1},
 \end{equation}
where $\Pi$ is the permutation matrix, eq.(\ref{70}) is
transformed to
 \be\label{72}
 R_{23}(z_2,z_3)R_{13}(z_1,z_3)R_{12}(z_1,z_2)=
 R_{12}(z_1,z_2)R_{13}(z_1,z_3)R_{23}(z_2,z_3).
 \end{equation}
This is the spectral Yang-Baxter equation.

The Bethe ansatz solution exists, if the scattering matrix
satisfies (\ref{70}). In other words, the matrix $b$ in (\ref{66})
and $c$ in (\ref{67}) and (\ref{68}) is acceptable only if the
resulting $S$-matrices satisfy (\ref{70}). This is a very
restricted condition and needed for having the solvability.

The $S$-matrices (\ref{66}) and (\ref{67}) are exactly the ones
considered in \cite{17} and \cite{16}, respectively. Using the
fact that $S^{(1)}$ is a binomial of degree one with respect to
$z_1^{-1}=e^{-ip_1}$ and $S^{(2)}$ is of degree one with respect
to $z_2$, it can be shown that SYB equation (\ref{70}) for
$S^{(1)}$ and $S^{(2)}$ reduces to
 \be\label{73}
b_{23}[b_{23},b_{12}]=[b_{23},b_{12}]b_{12},
\end{equation}
and
 \be\label{74} c_{12}[c_{12},c_{23}]=[c_{12},c_{23}]c_{23},
\end{equation}
respectively \cite{16,17}. Note that although the above equations
are much simpler than eq.(\ref{70}), but they are very complicated
yet. In $p$-species, each one is an equality between two
$p^3\times p^3$ matrices which results a system of $p^6$ equations
to be solved for $p^4-p^2$ elements of $b$ (\rm or $c$), which may
or may not have solution (eq.(\ref{46}) and (\ref{50}) reduce the
number of independent elements of $b$ and $c$ to $p^4-p^2$). The
general properties of the solutions of eq.(\ref{73}) and
eq.(\ref{74}) have been discussed in \cite{17} and \cite{16},
respectively, which can be directly used here. In other words, for
every solution of eq.(\ref{74}), there exists a corresponding
solvable model which have been discussed in \cite{16}, i.e.
\bea\label{75}
 A_\alpha\emptyset &\rightarrow &\emptyset A_\alpha \ \ \ \ {\rm with
 \ rate}\ \ 1,\cr
  A_\alpha A_\beta &\rightarrow & A_\gamma A_\delta  \ \ \ \ {\rm with
  \ rate} \ \ c^{\alpha\beta}_{\gamma\delta}.
 \eea
and a type 2 model with reactions written in eq.(\ref{51}) (note
that the reactions (\ref{75}) are a subset of (\ref{51}) with
$D_L=0$). The same is true for solutions of (\ref{73}). They can
describe the following solvable model (discussed in \cite{17}):
\bea\label{76}
 A_\alpha\emptyset &\rightarrow &\emptyset A_\alpha \ \ \ \ {\rm with
 \ rate}\ \ 1,\cr
 A_{\a_0}\cdots A_{\a_n}\emptyset &\rightarrow &\emptyset A_{\g_0}\cdots A_{\g_n}  \ \ \ \ {\rm with
  \ rate} \ \ (b_{n-1,n}\cdots
  b_{0,1})^{\a_0\cdots\a_n}_{\g_0\cdots\g_n},
 \eea
and a type 1 model with reaction (\ref{48}) (again at $D_L=0$,
(\ref{48}) reduces to (\ref{76})).

The reasoning which leads the SYB equations of $S^{(1)}$ and
$S^{(2)}$ to (\ref{73}) and (\ref{74}) does not work for $S^{(3)}$
since it is not a binomial of degree one with respect to $z_1$ or
$z_2$, in fact it contains all powers of $z_1$ and $z_2$. So
obtaining the solutions of (\ref{70}) for $S^{(3)}$ is more
difficult than for $S^{(1)}$ and $S^{(2)}$, even in the simplest
case $p=2$. In $p=2$ case we encounter a system of 64 equations
that must be solved for 12 non-diagonal elements of $c$ (the
diagonal elements are determined by eq.(\ref{55})). The solution
must be momentum-independent (independent of $z_1$, $z_2$ and
$z_3$) and non-negative. We can not solve this equations generally
( taking all $c_{ij}\neq0$ ) by standard mathematical softwares
and therefore restrict ourselves to some specific cases. For
example taking
\begin{equation}
\label{77}
 c=\left(
 \begin{array}{cccc}
 c_{11} &0 & 0 & c_{14} \\
 c_{21} &1 &0 & c_{24} \\
 1-c_{11}-c_{21} & 0 &1 &1-c_{14}-c_{24}\\
 0 & 0 & 0 &0
 \end{array}
 \right) ,
 \end{equation}
or
\begin{equation}
\label{78}
 c=\left(
 \begin{array}{cccc}
 0 &0 & 0 & 0 \\
 c_{21} &1 &0 & c_{24} \\
 c_{31} & 0 &1 &c_{34}\\
 1-c_{21}-c_{31} & 0 & 0 &1-c_{24}-c_{34}
 \end{array}
 \right) ,
 \end{equation}
which are the four-parameters cases, one obtains two solutions
\begin{equation}
\label{79}
 c=\left(
 \begin{array}{cccc}
 0 &0 & 0 & 0 \\
 D_R &1 &0 & D_L \\
 D_L & 0 &1 &D_R\\
 0 & 0 & 0 & 0
 \end{array}
 \right) ,
 \end{equation}
and one with $D_L \leftrightarrow D_R$. Taking
\begin{equation}
\label{80}
 c=\left(
 \begin{array}{cccc}
 1-c_{41} &1-c_{42} & 1-c_{43} & 0 \\
 0 &0 &0 & 0 \\
 0 & 0 &0 &0\\
 c_{41} & c_{42} & c_{43} & 1
 \end{array}
 \right) ,
 \end{equation}
or
\begin{equation}
\label{81}
 c=\left(
 \begin{array}{cccc}
 1 &1-c_{42} & 1-c_{43} & 1-c_{44} \\
 0 &0 &0 & 0 \\
 0 & 0 &0 &0\\
 0 & c_{42} & c_{43} & c_{44}
 \end{array}
 \right),
 \end{equation}
as some three-parameters cases, we find four solutions
\begin{equation}
\label{82}
 c=\left(
 \begin{array}{cccc}
 1 &1-c_{42} & 1-c_{43} & 0 \\
 0 &0 &0 & 0 \\
 0 & 0 &0 &0\\
 0 & c_{42} & c_{43} & 1
 \end{array}
 \right) ,
 \end{equation}
where each of $c_{42}$  and $ c_{43}$ are either $D_R$  or $D_L$.
Taking $A_1 \equiv A$ and $A_2 \equiv B$, the interactions
introduced by (\ref{79}), for instance, are
  \bea\label{83}
 A\emptyset &\stackrel{D_R}\rightarrow & \emptyset A \cr
 B\emptyset  &\stackrel{D_R}\rightarrow &\emptyset B \cr
 \emptyset A&\stackrel{D_L}\rightarrow & A \emptyset \cr
\emptyset B&\stackrel{D_L}\rightarrow & B \emptyset \cr
 AA &\stackrel{D_R}\rightarrow & AB \cr
 BB &\stackrel{D_L}\rightarrow & AB \cr
 AA &\stackrel{D_L}\rightarrow & BA \cr
 BB &\stackrel{D_R}\rightarrow & BA.
 \eea
 The model built on the reactions (\ref{83}) is integrable.

 Assuming that the solvability condition (\ref{70}) is satisfied,
 it is easy to see that the conditional probability (the
 propagator) is
\begin{equation}\label{84}
U({\mathbf{x}};t|{\mathbf{y}};0)=\int\frac
 {d^Np}{(2\pi )^N}e^{-E_Nt}e^{-i\mathbf{p.y}}
 \sum_\sigma {\mathbf{A}}_\sigma e^{i\sigma
({\mathbf{p}}).{\mathbf{x}}},
\end{equation}
where the integration region for each $p_i$ is $[0,2\pi]$ and
$A_1=1$. The singularity in $A_\sigma$ is removed by setting $p_j
\rightarrow p_j+i \varepsilon $, where one should consider the
limit $\varepsilon\rightarrow0^+$. Using this propagator, one can
write the probability at the time $t$  in terms of the initial
value of probability:
\begin{equation}\label{85}
|P({\mathbf{x}};t)\rangle=\sum_{y}U({\mathbf{x}};t|{\mathbf{y}};0)|P({\mathbf{y}};0)\rangle.
\end{equation}
Note that although $S^{(1)}$ and $S^{(2)}$ are similar to ones
considered in \cite{17} and \cite{16}, but the propagators
$U^{(1)} $and $U^{(2)}$ are different since the energy spectrum of
our models differs from those considered there. In $D_L=0$, our
results must coincide with those obtained in \cite{16,17}.

For the two-particle sector, there is only one matrix in the
expression of $U^{(i)}$s ($b$ in $U^{(1)}$ and $c$ in $U^{(2)}$
and $U^{(3)}$). So it can be treat as a $c$-number. Using
calculation similar to what has been done in \cite{16}-\cite{18},
one arrives at:
 \bea\label{86}
 U^{(1)}(x_1,x_2;t|y_1,y_2;0)& = &
 e^{-2t}\sum\limits_{n,m=0}^\infty \{ \frac{ D^n_LD_R^{x_1-y_1+n}t^{x_1-y_1+2n}}{n!(x_1-y_1+n)!}
\frac{ D_L^mD_R^{x_2-y_2+m}t^{x_2-y_2+2m}}{m!(x_2-y_2+m)!}\cr
&&+\sum\limits_{l=0}^\infty \frac{
 D_L^{n}D_R^{x_2-y_1+n}t^{x_2-y_1+2n}}{n!(x_2-y_1+n)!}\frac{D_L^{m}D_R^{x_1-y_2-l+m}t^{x_1-y_2-l+2m}}
  {m!(x_1-y_2-l+m)!}\cr
  && \ \ \ \times b^{l}[-1+\frac{{x_2-y_1+n}}{D_Rt}b] \}  ,
   \eea
and

 \bea\label{87}
 U^{(2)}(x_1,x_2;t|y_1,y_2;0)& = &
 e^{-2t}\sum\limits_{n,m=0}^\infty \{ \frac{ D_L^{n}D_R^{x_1-y_1+n}t^{x_1-y_1+2n}}{n!(x_1-y_1+n)!}\frac{D_L^{m}D_R^{x_2-y_2+m}t^{x_2-y_2+2m}}
  {m!(x_2-y_2+m)!}\cr
  &&+\sum\limits_{l=0}^\infty \frac{ D_L^{n}D_R^{x_2-y_1+l+n}t^{x_2-y_1+l+2n}}{n!(x_2-y_1+l+n)!}\frac{D_L^{m}D_R^{x_1-y_2+m}t^{x_1-y_2+2m}}
  {m!(x_1-y_2+m)!}\cr &&  \ \ \ \times c^{l}[-1+\frac{D_Rt}{{x_1-y_2+m+1}}c] \}.
   \eea
Similarly one can obtain a more lengthy expression for $U^{(3)}$.
Note that at $D_L=0$, eqs.(\ref{86}) and (\ref{87}) lead
eqs.(\ref{38}) of \cite{17} and (\ref{30}) of \cite{16},
respectively.

To investigate the large-time behaviours of the probabilities
$U^{(1)}$, $U^{(2)}$, and $U^{(3)}$, it is useful to decompose the
vector spaces on which $b$ (in type 1 model) and $c$ (in types 2
and 3 models) act, in two subspaces invariant under the action of
$b(c)$: the first subspace corresponding to eigenvalues with
modulus one, and another invariant subspace. For types 1 and 2
models with conditions (\ref{46}) and (\ref{50}), as all the
elements of matrix $b(c)$ are non-negative, the second subspace
corresponds to eigenvalues with modulus less than one. By focusing
on type 1 model, this decomposition can be done by introducing two
projectors $Q$ and $R$, satisfying
 \bea\label{88}
  Q+R&=&1,\cr QR=RQ&=&0, \cr [b,Q]=[b,R]&=&0.
   \eea
$Q$ projects on the first subspace and $R$ projects on the second.
Following \cite{16}, we multiply $U^{(1)}$ by $Q+R=1$:
  \begin{equation}\label{89}
 U^{(1)}({\bf x};t|{\bf y};0)=U^{(1)}Q+U^{(1)}R.
 \end{equation}
In the terms multiplied by $R$, one can treat $b$ as a number with
modulus different from one. So the integrand in (\ref{84}) is
non-singular at points $p_j=0$, which have the main contributions
at large times. Putting $p_j=0$, we have $S^{(1)}\approx -1$ and
${\mathbf A}_{\sigma}\approx (-1)^{[\sigma]}$, and eq.(\ref{84})
results
 \bea\label{90}
 {\rm the\ second\ term\ of\ }U^{(1)}=\frac{1}{2\pi
 t}\{&&
  e^{-\left\{ \left[ x_1-y_1-(D_R-D_L)t\right]^2+ \left[ x_2-y_2-(D_R-D_L)t\right]^2
 \right\}/(2t)} \cr &-&
  e^{-\left\{ \left[ x_1-y_2-(D_R-D_L)t\right]^2+ \left[ x_2-y_1-(D_R-D_L)t\right]^2
 \right\}/(2t) } \} R, \ \ \ t\rightarrow\infty ,\cr &&
 \eea
which is independent of $b$. So at large time, the second term of
$U^{(1)}$ tends to zero faster than $t^{-1}$ and the leading term
in $U^{(1)}$, which is order $t^{-1}$, does not involve the second
term.

If the only eigenvalue of $b$ with modulus 1 is 1, then $bQ=Q$ and
$U^{(1)}$ has a simple behaviour at $t\rightarrow\infty$:
 \bea\label{91}
 U^{(1)}(x_1,x_2;t|y_1,y_2;0)& = &
 e^{-2t}\sum\limits_{n,m=0}^\infty \{ \frac{ D^n_LD_R^{x_1-y_1+n}t^{x_1-y_1+2n}}{n!(x_1-y_1+n)!}
\frac{ D_L^mD_R^{x_2-y_2+m}t^{x_2-y_2+2m}}{m!(x_2-y_2+m)!}\cr
&&+\sum\limits_{l=0}^\infty \frac{
 D_L^{n}D_R^{x_2-y_1+n}t^{x_2-y_1+2n}}{n!(x_2-y_1+n)!}\frac{D_L^{m}D_R^{x_1-y_2-l+m}t^{x_1-y_2-l+2m}}
  {m!(x_1-y_2-l+m)!}\cr
  && \ \ \ \times [-1+\frac{{x_2-y_1+n}}{D_Rt}] \} Q  .
   \eea
This is simply the propagator of a single-species model with
diffusions to right and left and drop-push to right (i.e. the
$\lambda =0$ case of the reactions studied in \cite{19} and
\cite{30}), multiplied by $Q$. In fact eq.(\ref{91}) is $\lambda
=0$ case of eq.(30) of \cite{19}, times $Q$.

For $U^{(2)}$, the same decomposition leads the eq.(\ref{90}) for
its second term and in the case $cQ=Q$, $U^{(2)}$ tends to
(\ref{87}), with $c=1$, times $Q$, at $t\rightarrow\infty$. The
resulting one-species model is $\mu =0$ case of the reactions
studied in \cite{19} and \cite{30}. For $U^{(3)}$, we again find
(\ref{90}) and the one-species partially asymmetric simple
exclusion process of \cite{13}.

 {\bf Acknowledgement:}  We would like to thank the
research council of
 the University of Tehran for partial financial support.

\end{document}